\documentclass{article}
\usepackage{PRIMEarxiv}
\usepackage[utf8]{inputenc} 
\usepackage[T1]{fontenc}    
\usepackage{url}            
\usepackage{makecell}
\usepackage{amsfonts}       
\usepackage{nicefrac}       
\usepackage{microtype}      
\usepackage{lipsum}
\usepackage{fancyhdr}       
\usepackage{graphicx}       
\graphicspath{{media/}}     
\usepackage{xcolor}
\usepackage[acronym]{glossaries}
\usepackage{siunitx}
\usepackage{appendix}
\usepackage{multirow}
\usepackage{caption}
\usepackage{subcaption}
\usepackage{hyperref}
\usepackage{tabularray}
\usepackage{amssymb}
\usepackage{pifont}
\newcommand{\cmark}{\ding{51}}%
\newcommand{\xmark}{\ding{55}}%
\usepackage[font=small]{caption}
\usepackage[hang,flushmargin]{footmisc}
\usepackage[flushleft]{threeparttable}
\usepackage{cleveref}
\usepackage{multicol}
\usepackage{float}
\usepackage{booktabs} 
\crefname{figure}{Fig.}{Figs.}

\usepackage[
backend=biber,
style=ieee,
citestyle=numeric-comp,
maxbibnames=8,
minbibnames=4,
sorting=none,
url=false,
doi=true,
isbn=false,
uniquename=false,
alldates=year,
clearlang=true,]{biblatex}

\AtEveryBibitem{\clearlist{publisher}}
\AtEveryBibitem{\clearfield{note}}
\AtEveryBibitem{\clearlist{institution}}
\AtEveryBibitem{\clearlist{location}}
\AtEveryBibitem{\clearlist{language}}
\AtEveryBibitem{\clearname{editor}}

\usepackage{doi}

\DeclareSIUnit\mac{MAC}

\DeclareFieldFormat[article,unpublished,misc,preprint,inproceedings]{title}{``#1''}

\let\oldcite\cite
\renewcommand{\cite}[1]{{\mbox{\oldcite{#1}}}}

\renewcommand{\arraystretch}{1.3} 

\title{Fused-MemBrain: a spiking processor combining CMOS and self-assembled memristive networks}

\author{
  \small Davide Cipollini\textsuperscript{*,1,3}, Hugh Greatorex\textsuperscript{*2,3}, Michele Mastella\textsuperscript{2,3}, Elisabetta Chicca\textsuperscript{2,3}, Lambert Schomaker\textsuperscript{1,3}\\
  \small \textsuperscript{1}Bernoulli Institute for Mathematics, Computer Science and Artificial Intelligence, University of Groningen, Netherlands\\
  \small \textsuperscript{2}Bio-Inspired Circuits and Systems (BICS) Lab, Zernike Institute for Advanced Materials, University of Groningen, Netherlands\\
  \small \textsuperscript{3}Cognigron - Groningen Cognitive Systems and Materials Center, University of Groningen, Netherlands\\
  \small \textsuperscript{*}Corresponding authors: \texttt{d.cipollini@rug.nl, h.r.greatorex@rug.nl}
}

\PassOptionsToPackage{acronym}{glossaries}

\glsdisablehyper

\newcommand*\myglsentry[1]{%
   \glsentrylong{#1}%
}

\newacronym[longplural={Frames per Second}]{fpsLabel}{FPS}{Frame per Second}
\newacronym[longplural={Tables of Contents}]{tocLabel}{TOC}{Table of Content}
\newacronym{act}{ACT}{Asynchronous Circuit Toolkit}
\newacronym{adc}{ADC}{Analog Digital Converter}
\newacronym{aer}{AER}{Address Event Representation}
\newacronym{afe}{AFE}{Analog FrontEnd}
\newacronym{ai}{AI}{Artificial Intelligence}
\newacronym{ampa}{AMPA}{$\alpha$-amino-3-hydroxy-5-methyl-4-isoxazolepropionic acid}
\newacronym{ams}{AMS}{Analog Mixed-Signal}
\newacronym{api}{API}{Application Programming Interface}
\newacronym{bap}{BAP}{back-propagating action potential}
\newacronym{bcall}{BCaLL}{Bistable Calcium-based Local Learning}
\newacronym{bd}{BD}{Bundled Data}
\newacronym{beol}{BEOL}{Back-End Of Line}
\newacronym{bjt}{BJT}{Bipolar Junction Transistor}
\newacronym{bnc}{BNC}{Bayonet-Neill-Concelman}
\newacronym{bptt}{BPTT}{Back Propagation Through Time}
\newacronym{bp}{BP}{Back Propagation}
\newacronym{camp}{cAMP}{cyclic Adenosine Mono Phosphate}
\newacronym{cam}{CAM}{Content Addressable Memory}
\newacronym{cco}{CCO}{Current Controlled Oscillator}
\newacronym{cc-by}{CC-BY}{Creative Commons Attribution}
\newacronym{chp}{CHP}{Communicating Hardware Processes}
\newacronym{cl}{CL}{Closed-Loop}
\newacronym{cm}{CM}{Continuum Mechanics}
\newacronym{cpu}{CPU}{Central Process Unit}
\newacronym{csv}{CSV}{Comma-Separated Values}
\newacronym{ctat}{CTAT}{Complementary To Absolute Temperature}
\newacronym{cuba}{CUBA}{Current Based}
\newacronym{dac}{DAC}{Digital to Analog Converter}
\newacronym{dcn}{DCN}{Dorsal Column Nuclei}
\newacronym{dc}{DC}{Direct Current}
\newacronym{dpi}{DPI}{Differential Pair Integrator}
\newacronym{dpss}{DPSS}{Dendritic Prediction of Somatic Spiking}
\newacronym{drg}{DRG}{Dorsal Root Ganglion}
\newacronym{dr}{DR}{Dual Rail}
\newacronym{dsnn}{D-SNN}{Deep Spiking Neural Network}
\newacronym{dsp}{DSP}{Digital Signal Processing}
\newacronym{ds}{DS}{Delay Sensitive}
\newacronym{dvs}{DVS}{Dynamic Vision Sensor}
\newacronym{eda}{EDA}{Electronic Design Automation}
\newacronym{elm}{ELM}{Extreme Learning Machine}
\newacronym{esn}{ESN}{Echo State Network}
\newacronym{fac}{FAC}{Facilitatory Trace}
\newacronym{fecap}{FeCap}{Ferroelectric Capacitor}
\newacronym{fet}{FET}{Field Effect Transistor}
\newacronym{fft}{FFT}{Fast Fourier Transform}
\newacronym{fifo}{FIFO}{First In First Out Register}
\newacronym{fi}{FI}{Frequency vs. Current}
\newacronym{fm}{FM}{Frequency Modulated}
\newacronym{fo}{FO}{First Order}
\newacronym{fpga}{FPGA}{Field Programmable Gate Array}
\newacronym{fsm}{FSM}{Finite State Machine}
\newacronym{ftj}{FTJ}{Ferroelectric Tunnel Junction}
\newacronym{ft}{FT}{Fourier transform}
\newacronym{gmm}{GMM}{Gaussian mixture models}
\newacronym{gpu}{GPU}{Graphic Process Unit}
\newacronym{hp}{HP}{High Pass Filter}
\newacronym{i2c}{I$^2$C}{Inter Integrated Circuit}
\newacronym{ic}{IC}{Integrated Circuit}
\newacronym{if}{IF}{Integrate and Fire}
\newacronym{imc}{IMC}{In-Memory Computing}
\newacronym{iot}{IoT}{Internet of Things}
\newacronym{io}{I/O}{Input and Output}
\newacronym{ip}{IP}{Intellectual Property Macro}
\newacronym{irh}{IRH}{Instantaneous Rate Histogram}
\newacronym{isi}{ISI}{Interspike Interval}
\newacronym{iwta}{I-WTA}{Inverted Winner-Take-All}
\newacronym{knn}{K-NN}{K-Nearest Neighbors}
\newacronym{lcadc}{LC-ADC}{Level Crossing Analog Digital Converter}
\newacronym{lda}{LDA}{Linear Discriminant Analysis}
\newacronym{lif}{LIF}{Leaky \myglsentry{if}}
\newacronym{llm}{LLM}{Large Language Model}
\newacronym{lpf}{LPF}{Low Pass Filter}
\newacronym{lstm}{LSTM}{Long Short-Term Memory}
\newacronym{ltp}{LTP}{Long Term Plasticity}
\newacronym{lut}{LUT}{Look Up Table}
\newacronym{lvds}{LVDS}{Low Voltage Differential Signalling}
\newacronym{mac}{MAC}{Multiply Accumulate}
\newacronym{mems}{MEMS}{Micro Electro-Mechanical System}
\newacronym{mim}{MIM}{Metal Insulator Metal}
\newacronym{mlm}{MLM}{Multi Layer Mask}
\newacronym{mlp}{MLP}{Multi Layer Perceptron}
\newacronym{ml}{ML}{Machine Learning}
\newacronym{mnist}{MNIST}{modified National Institute of Standards and Technology database}
\newacronym{mom}{MOM}{Metal Oxide Metal}
\newacronym{mos}{MOS}{Metal Oxide Semiconductor}
\newacronym{mtj}{MTJ}{Magnetic Tunnel Junction}
\newacronym{mvm}{MVM}{Matrix-Vector Multiplication}
\newacronym{nas}{NAS}{Neuromorphic Auditory Sensor}
\newacronym{nda}{NDA}{Non Disclosure Agreement}
\newacronym{nmda}{NMDA}{N-methyl-D-aspartate}
\newacronym{nni}{NNI}{Neural Network Intelligence}
\newacronym{nn}{NN}{Neural Network}
\newacronym{noc}{NoC}{Network on Chip}
\newacronym{nvm}{NVM}{Non-Volatile Memory}
\newacronym{oa}{OA}{OpenAccess}
\newacronym{opamp}{OPAMP}{Operational Amplifier}
\newacronym{ota}{OTA}{Operational Transconductance Amplifier}
\newacronym{pca}{PCA}{Principal Component Analysis}
\newacronym{pcb}{PCB}{Printed Circuit Board}
\newacronym{pcfb}{PCFB}{Pre-Charge Full Buffer}
\newacronym{pchb}{PCHB}{Pre-Charge Half Buffer}
\newacronym{pcm}{PCM}{Phase Change Material}
\newacronym{pc}{PC}{Pacini}
\newacronym{pdk}{PDK}{Process Development Kit}
\newacronym{pd}{PD}{Phase Detector}
\newacronym{pll}{PLL}{Phase-Locked Loop}
\newacronym{posfet}{POS-FET}{Piezoelectric Oxide Semiconductor Field Effect Transistor}
\newacronym{prs}{PRS}{Production Rule Set}
\newacronym{pr}{P\&R}{Place and Route}
\newacronym{psc}{PSC}{Post Synaptic Current}
\newacronym{ptat}{PTAT}{Proportional To Absolute Temperature}
\newacronym{pv}{PV}{Parietal Ventral Area}
\newacronym{qdi}{QDI}{Quasi Delay Insensitive}
\newacronym{qfp}{QFP}{Quad-Flat Package}
\newacronym{ra1}{RA}{Rapid-Adapting I}
\newacronym{ra2}{RA2}{Rapid-Adapting II}
\newacronym{rafr}{RaFr}{Radio Frequency}
\newacronym{ram}{RAM}{Random Access Memory}
\newacronym{rbssg}{RBSSG}{Reverse Bitwise Synthetic Spike Generator}
\newacronym{rf}{RF}{Receptive Field}
\newacronym{rl}{RL}{Reinforcement Learning}
\newacronym{rnn}{RNN}{Recurrent Neural Network}
\newacronym{roi}{ROI}{Region Of Interest}
\newacronym{rpe}{RPE}{Reward Prediction Error}
\newacronym{s1}{S1}{Somatosensory Primary Cortex}
\newacronym{s2}{S2}{Somatosensory Secondary Cortex}
\newacronym{sa1}{SA1}{Slow-Adapting I}
\newacronym{sa2}{SA2}{Slow-Adapting II}
\newacronym{sdsp}{SDSP}{Spike-Driven Synaptic Plasticity}
\newacronym{sfd}{SFD}{Spike Frequency Divider}
\newacronym{shf}{S-HF}{Spike-based Hold \& Fire}
\newacronym{sig}{S-IG}{Spike-based Integrate \& Fire}
\newacronym{simd}{SIMD}{Single Instruction, Multiple Data}
\newacronym{slpf}{S-LPF}{Spike-based Low Pass Filter}
\newacronym{smd}{SMD}{Surface Mount Device}
\newacronym{smu}{SMU}{Source Measurement Unit}
\newacronym{soc}{SoC}{System on Chip}
\newacronym{sota}{SOTA}{State-Of-The-Art}
\newacronym{so}{SO}{Second Order}
\newacronym{spice}{SPICE}{Simulation Program with Integrated Circuit Emphasis}
\newacronym{spi}{SPI}{Serial Peripheral Interface}
\newacronym{spll}{sPLL}{Spiking Phase-Locked Loop}
\newacronym{src}{SRC}{Sparse Representation Classifier}
\newacronym{srdp}{SRDP}{Spike-Rate-Dependent Plasticity}
\newacronym{srm}{SRM}{Simple Response Model}
\newacronym{stdp}{STDP}{Spike Timing Dependent Plasticity}
\newacronym{stp}{STP}{Short Term Plasticity}
\newacronym{svm}{SVM}{Support Vector Machine}
\newacronym{syn}{SYN}{Synaptic Trace}
\newacronym{tc}{TC}{Temporal Contrast}
\newacronym{tde}{TDE}{Time Difference Encoder}
\newacronym{tpu}{TPU}{Tensor Processing Unit}
\newacronym{trg}{TRG}{Trigger Trace}
\newacronym{ttfs}{TTFS}{Time to first spike}
\newacronym{uc}{$\mu$C}{Microcontroller}
\newacronym{vco}{VCO}{Voltage Controlled Oscillator}
\newacronym{vpd}{VP-d}{Victor-Purpura Distance}
\newacronym{wta}{WTA}{Winner Take All}

\newacronym{alif}{ALIF}{Adaptive \myglsentry{lif}}
\newacronym{ann}{ANN}{Artificial \myglsentry{nn}}
\newacronym{asic}{ASIC}{Application Specific \myglsentry{ic}}
\newacronym{cmos}{CMOS}{Complementary \myglsentry{mos}}
\newacronym{cnn}{CNN}{Convolutional \myglsentry{nn}}
\newacronym{cstdp}{C-STDP}{Calcium \myglsentry{stdp}}
\newacronym{ddpi}{DDPI}{Double \myglsentry{dpi}}
\newacronym{dnn}{DNN}{Deep \myglsentry{nn}}
\newacronym{eai}{Edge-AI}{Edge \myglsentry{ai}}
\newacronym{enn}{ENN}{event-based \myglsentry{nn}}
\newacronym{epsc}{EPSC}{Excitatory \myglsentry{psc}}
\newacronym{exlif}{ExLIF}{Exponential \myglsentry{lif}}
\newacronym{fefet}{FeFET}{Ferroelectric \myglsentry{fet}}
\newacronym{gpgpu}{GPGPU}{General-Purpose computing on \myglsentry{gpu}}
\newacronym{ipsc}{IPSC}{Inhibitory \myglsentry{psc}}
\newacronym{moscap}{MOSCAP}{\myglsentry{mos} Capacitor}
\newacronym{mosfet}{MOSFET}{\myglsentry{mos} \myglsentry{fet}}
\newacronym{nmnist}{N-MNIST}{Neuromorphic \myglsentry{mnist}}
\newacronym{nmos}{nMOS}{n-type \myglsentry{mos}}
\newacronym{oxram}{OxRAM}{Oxide-based resistive \myglsentry{ram}}
\newacronym{pka}{PKA}{\myglsentry{camp}-depended Protein Kinase}
\newacronym{pmos}{pMOS}{p-type \myglsentry{mos}}
\newacronym{rram}{RRAM}{Resistive \myglsentry{ram}}
\newacronym{sadc}{sADC}{spiking \myglsentry{adc}}
\newacronym{scnn}{sCNN}{spiking \myglsentry{cnn}}
\newacronym{snn}{SNN}{Spiking \myglsentry{nn}}
\newacronym{sodpi}{SoDPI}{Second-order \myglsentry{dpi}}
\newacronym{sram}{SRAM}{Static \myglsentry{ram}}
\newacronym{tstdp}{T-STDP}{Triplet \myglsentry{stdp}}

\newacronym{adexlif}{AdExLIF}{Adaptive Exponential Leaky Integrate-and-Fire}
\newacronym{nfet}{n-FET}{\myglsentry{nmos} \myglsentry{fet}}
\newacronym{pfet}{p-FET}{\myglsentry{pmos} \myglsentry{fet}}

\newacronym{sma}{SMA}{SubMiniature version A}
\newacronym{wrota}{WR-OTA}{Wide Range \gls{ota}}
\newacronym{sstdp}{S-STDP}{Stochastic \myglsentry{stdp}}
\newacronym{neuop}{NeuOp}{Neuron spike Operation}
\newacronym{synop}{SynOp}{Synaptic Operation}
\newacronym{fdsoi}{FDSOI}{Fully Depleted Silicon On Insulator}
\newacronym{qif}{QIF}{Quadratic \myglsentry{if}}
\newacronym{ff}{FF}{edge triggered Flip-Flop}
\newacronym{lrs}{LRS}{Low Resistance State}
\newacronym{hrs}{HRS}{High Resistance State}
\newacronym{mvna}{MVNA}{Modified Voltage Nodal Analysis}
\newacronym{lsm}{LSM}{Liquid State Machines}

\makeglossaries

\begin{document}
\maketitle

\begin{abstract}

In an era characterized by the rapid growth of data processing, developing new and efficient data processing technologies has become a priority.
We address this by proposing a novel type of neuromorphic technology we call Fused-MemBrain.
Our proposal is inspired by Golgi's theory modeling the brain as a syncytial continuum, in contrast to Cajal's theory of neurons and synapses being discrete elements.
While Cajal's theory has long been the dominant and experimentally validated view of the nervous system, recent discoveries showed that a species of marine invertebrate (ctenophore \emph{Mnemiopsis leidyi}) may be better described by Golgi's theory.
The core idea is to develop hardware that functions analogously to a syncytial network, exploiting self-assembled memristive systems and combining them with CMOS technologies, interfacing with the silicon back-end-of-line. 
In this way, a memristive self-assembled material can cheaply and efficiently replace the synaptic connections between CMOS neuron implementations in neuromorphic hardware, enhancing the capability of massively parallel computation. 
The fusion of CMOS circuits with a memristive ``plexus'' allows information transfer without requiring engineered synapses, which typically consume significant area.
As the first step toward this ambitious goal, we present a simulation of a memristive network interfaced with spiking neural networks. 
Additionally, we describe the potential benefits of such a system, along with key technical aspects it should incorporate.

\end{abstract}


\vspace{+10pt}

\begin{multicols}{2}

\section{Introduction}

\begin{figure*}
    \centering
    \includegraphics[width=\linewidth]{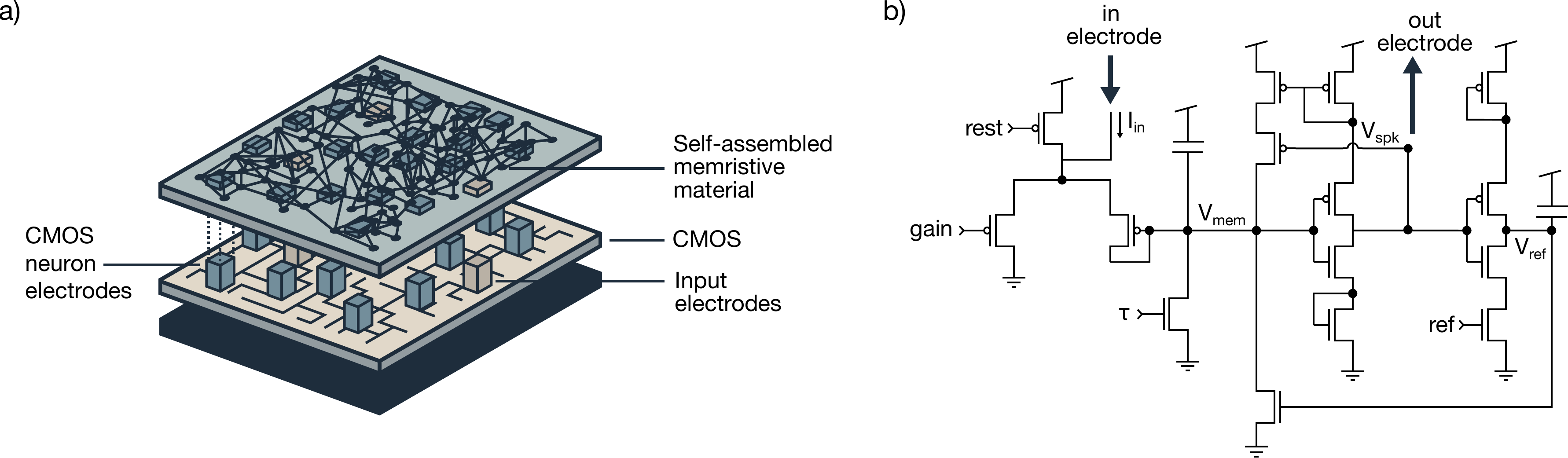}
    \caption{\textbf{The proposed Fused-MemBrain processor.} 
    \textbf{a)} A schematic depiction of the proposed hardware. 
    A CMOS layer contains all of the neuron circuitry and configuration. 
    On top, a self-assembled memristive material is deposited. 
    The interfacing of the two layers is facilitated by electrodes protruding from the uppermost layers of \gls{cmos}.
    \textbf{b)} an exemplary neuron circuit adapted from~\cite{Chicca2014}. 
    The neuron integrates input currents from the plexus through one of its two electrodes (in electrode).
    Once the neuron's membrane voltage crosses its firing threshold, it generates and transmits a voltage pulse (spike). 
    This pulse can be fed back into the plexus via another electrode (out electrode), stimulating recurrent network activity.}
    \label{fig:fusedmembrain}
\end{figure*}

Artificial information processing has already become one of the main contributors to the current climate and energy crisis~\cite{patterson_carbon_2021}. 
Vast amounts of data are continuously generated and fed into our computing machines.
Moreover, sophisticated AI tools, such as Large Language Models, are currently over-utilized to handle even the simplest tasks, largely due to their public and nearly limitless availability~\cite{patterson_carbon_2021}.
While chip manufacturing companies strive to find new solutions to continue the miniaturization of integrated circuits, the physical limitations are dawning upon them~\cite{waldrop2016chips}.
With this on the horizon, emerging and specialized computing schemes are poised to become viable solutions, particularly when integrated with developing devices and materials.

Neuromorphic engineering~\cite{Mead2020,indiveri2011neuromorphic} takes inspiration from the remarkable energy efficiency of the brain~\cite{Markovi2020} and leverages the physics of devices such as transistors to mimic the dynamical behavior of neural circuits.
Often, the field grapples with the challenge of achieving reliable computation from unreliable computational elements~\cite{Cotteret2024_distributed}, in contrast to modern digital signal processing circuits designed to be robust to noise.
In stark contrast, brain function is underpinned by billions of neurons and trillions of synapses and dendrites communicating via analog (potentials) and digital signals (spikes).
The biological computing paradigm is inherently stochastic, redundant, and resilient to noise and individual defects, standing in bold relief against the backdrop of modern digital computing.
Neuromorphic computing and engineering endeavor to remove, or at least minimize, the physical distance between CPU and memory, widely referred to as the infamous von Neumann bottleneck.
In this way, such hardware functions as a \textit{network-that-does} as much as the brain is, enabling higher bandwidth and significantly reduced energy consumption~\cite{McKee2004}.

The brain possesses an in-homogeneous topological organization~\cite{bullmore2009complex,barabasi1999emergence} in both its structural and functional wiring. 
This has a profound impact on its activity and its capabilities, so much so that abnormal structures lead to neuropsychiatric pathologies such as schizophrenia~\cite{micheloyannis2006small,liu2008disrupted}, epilepsy~\cite{kramer2008emergent}, and attention-deficit/hyperactivity disorder\cite{wang2009altered}.
Moreover, the brain functions as a spatial network~\cite{barthelemy2011spatial}, with nodes positioned in physical space and a topological structure shaped by physical constraints and resource limitations~\cite{zhang2024geometric}.

Many of these properties are not realized when cross-bar arrays of analog memristive memories are used to implement efficient in-memory computation. 
Cross-bar arrays operate to accelerate the matrix-vector multiplications at the foundation of artificial neural network algorithms, nevertheless, they become inefficient when recurrent dynamics are imposed, especially when matrices reflecting the coupling between elements are sparse off-diagonal.
The realization of specialized hardware for small-world connectivity has been addressed~\cite{dalgaty2024mosaic} requiring the shrewd top-down design of pairwise memristive synapses given specific properties of the wiring topology. 
Nevertheless, the bottom-up realization of self-assembled memristive materials~\cite{Vahl_2024} may allow the efficient, reconfigurable, and almost designless realization of multielectrode devices with the significant advantage of reducing the area the \gls{cmos}-synapses consume in hardware.

We propose making use of the memristive disordered topology of in-organic materials to implement the coupling, and conventional analog and digital electronics to implement the activity of the nodes.
We note that previous studies in the literature have demonstrated high-density multielectrode arrays integrated with disordered wetware with \textit{in-silico} computing~\cite{kagan2022vitro,isomura2018vitro,isomura2015cultured}.
\textit{In vitro} neurons embedded in high-density electrodes are already under consideration to harness the computational power of biological systems, and the self-adaption of their neural response to the environment upon free energy minimization has been demonstrated~\cite{friston2010free}. 
We present a similar multi-electrode system exploiting inorganic materials, provide an easy-to-use simulator for experimentation, and examine the potential implications of the proposed hardware.

Therefore the main contribution of this work is twofold:  

\begin{enumerate}
    \item The proposal for a spiking system equivalent to a syncytial network on hardware.
    \item The development of a compact simulator complementary to this proposal.
\end{enumerate}

\section{Fused network and the biological neuron doctrine}

One of the most significant technical advances in the investigation of the nervous system was the so-called ``black reaction'' method, invented by Camillo Golgi in the 19th century~\cite{Golgi1885}. 
With this new technique, Golgi could stain a small percentage of the elements within a block of nervous tissue.
This enabled the visualization of entire nerve cells, including their dendritic arborization and axons, using a microscope~~\cite{Glickstein2006}.
As a result of his experimental work, Golgi proposed his theory, suggesting that the nervous system was a \textit{syncytial continuum}.
According to his hypothesis, nerve cells were fused into a single giant membrane, forming a \textit{plexus} that served as the primary integrative structure of the nervous system.
Around the same time, another celebrated scientist, Santiago Ram\'{o}n y Cajal, adopted the black-reaction technique that Golgi pioneered. 
Opposing the fused network theory of his contemporary, Cajal hypothesized that the nervous system was composed of discrete elements, called neurons, that may interface with one another but never fuse.
Although both Golgi and Cajal won the Nobel Prize in Physiology or Medicine in 1906, the advent of electron microscopy, which enabled the discovery of synaptic connections between individual neurons~\cite{Palade54,Gray59}, ultimately led to the rejection of Golgi's theory in favor of the neuron doctrine proposed by Cajal.

However, there is some increasing evidence that for some jellyfish, such as the ctenophore \emph{Mnemiopsis leidyi}, nerve-net neurons connect through a syncytial continuum: a wide network extending from only one soma~\cite{Burkhardt2023}. 
This type of architecture may be appropriate if its main function is to display sensitivity for the relevant dynamics at the input (sensor) and have the required dynamical properties towards the output (actuator), i.e., general locomotion for foraging/oxygen intake and predator evasion, in the case of jellyfish, with its typical sensory reactivity and contractile reactions~\cite{Thoma2023}. 

In this work, we aim to propose and explore the possibilities of a comparable self-assembled, self-organized memristive spiking network in hardware, not as a replacement for other forms of neural processing, but as a complementary approach characterized by architectural simplicity while still being capable of handling the required non-linear dynamics of the associated real-world functions~\cite{Pallasdies2019,Yuan2014}.

\section{Fused-MemBrain: Hardware}

The proposed hardware is defined as combining two main components.
The first component is an ensemble of \gls{cmos} neurons, e.g. \gls{lif} equipped with two electrodes: one that sources currents and one that provides voltage. 
The \textit{in} electrode drains the electrical current from its surroundings, integrates it, and produces a voltage that the \textit{out} electrode applies back into the plexus. 
Such \gls{cmos} neurons are configurable and the external user can set their dynamical parameters to the requirements imposed by the task. 

The second component is the self-assembled memristive plexus deposited on the silicon back-end-of-line which couples the neuron electrodes. 
Thus, signals between coupled neurons are wave-like propagated \textit{in materio} through time and the physical space.

The emerging field of self-assembled neuromorphic materials~\cite{Sillin2013,Vahl_2024}, e.g. nanowire networks~\cite{Sillin2013,Zhu2023,Milano_2021_reservoir}, domain wall networks~\cite{rieck2022ferroelastic}, nanocluster-assembled devices~\cite{Profumo_2023,Mallinson2024}, disordered dopant-atom network~\cite{Chen2020}, disordered networks of nanodots~\cite{remacle1998networks,liu2021colloidal} present themselves as promising candidates to replace the synaptic circuitry.
The deposition of many of these materials, e.g. nanowires, nanodots, and nanocluster assemblies, being \gls{cmos}-compatible, can in principle be interfaced with the circuitry on the silicon back-end-of-line~\cite{greatorex2024texel}, providing an efficient and low-cost solution to signal routing across the ensemble of \gls{cmos} neurons. 
They can be used to reduce the area required for realizing synaptic circuitry, thus reducing the silicon real estate, while at the same time enhancing the capability of massively parallel computation for \textit{in-materio} computing.

\Cref{fig:fusedmembrain} schematizes the envisioned Fused-MemBrain hardware realization. 
The dynamic behavior of the self-assembled memristive material can, in principle, be exploited to implement short- or long-term plasticity between the neurons, which could realize self-healing and self-organizing recurrent dynamical systems.

\begin{figure}[H]
    \centering
    \includegraphics[width=\linewidth]{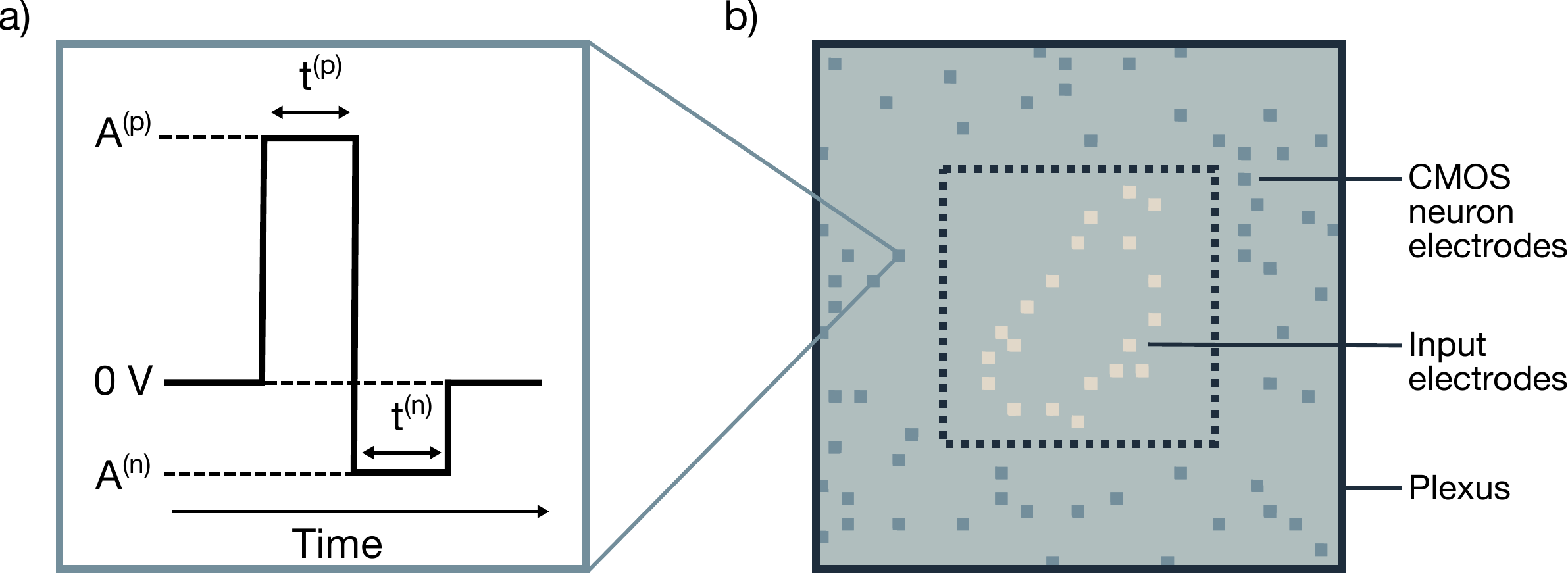}
    \caption{\textbf{CMOS neuron configuration.} 
    \textbf{a)} The spike-pulse shape is illustrated. 
    \textbf{b)} Input nodes are located in the central region of the plexus delimited by the square perimeter. 
    They inject the input data and drive the system out of equilibrium. 
    They are disposed to reflect the 20 higher-intensity coarse-grained pixels of a zero-digit sample from the MNIST dataset.}
    \label{fig:neuron_conf}
\end{figure}

\subsection{Configuration of the CMOS neurons}

\subsubsection{Spike shapes for synaptic and heterosynaptic plasticity}

The spike pulse, applied by the neurons onto the plexus, was engineered to be compatible with the electrical properties of the self-assembled material. 
Spikes are simplified as a positive square voltage followed by a negative square voltage as depicted in \mbox{\Cref{fig:neuron_conf}a}, mimicking the biological alternation between the spike and the refractory period. 
The amplitudes, $A^{(p)}$ and $A^{(n)}$, and the time widths of both the positive and negative square signals, $t^{(p)}$, $t^{(n)}$, are tunable parameters that determine the time-dependent behavior of the overall system.
The voltage difference between different neurons, generated by the shape of the spike (first positive, then negative), enables causal dependence between the spiking activity between two neurons in the same physical neighborhood. 
Let us take two neurons on the plexus, where one has entered the refractory period, $t^{(n)}$ and the other is currently applying the positive voltage, $A^{(p)}$, the voltage difference applied to the memristive plexus is the highest possible $\Delta V = A^{(p)} - A^{(n)}$ producing the enhancement of conductive pathways between these two electrodes resembling the synaptic plasticity.

\subsubsection{Electrode spatial distribution}

The Fused-Membrain hardware enables the deployment of multiple electrodes, unlike the simpler utilizing only two electrodes case as described in Ref.~\cite{Cipollini2023}, as a result, the voltage landscape over the plexus is significantly more complex. 
Since the synaptic plexus, rather than discrete connections between individual neuron pairs, realizes higher-order coupling, any change in coupling efficacy between one neuron pair will also influence the efficacy between other pairs, similar to heterosynaptic plasticity~\cite{lynch1977heterosynaptic}.
Therefore, given that there is no direct access to the connections between neurons and that the connections are of high order, we lose direct control of the synaptic connectivity and rely solely on the self-organization of the conductive memristive material. 
However, we retain control of the parameters governing the dynamics of the \gls{cmos} neurons along with their distribution on the plexus. See the last column of \Cref{tab:lif_parameters}. 
This is an alternative and tangible way to modify the spatiotemporal behavior of the plexus. 

In \Cref{fig:neuron_conf}b, we show an example distribution of the electrodes across the plexus.
We categorize the electrodes protruding into the memristive plexus as \textit{input} electrodes and the electrodes of the \gls{cmos} neurons. 
The input electrodes are located within the central region of the processor, indicated by the inner square in panel \Cref{fig:neuron_conf}b. 
These electrodes are used to inject the input signals and do not interact with the plexus after the presentation of the sample. 
As an example (\Cref{fig:neuron_conf}) the input electrodes are placed to reflect the location of the active pixels of a zero-digit sample from the MNIST dataset.  
The green squares indicate the distribution of the \gls{cmos} neurons.

Speculatively, an intriguing extension could involve introducing quantum dots into the input electrode region of the memristive plexus.
Thus, optical input signals encoding data could be efficiently converted into electrical signals propagating to the neuron-electrodes and initiating their activity~\cite{remacle1998networks,liu2021colloidal}.

\section{Fused-MemBrain: Simulator}

\begin{figure*}[hbt]
    \centering
    \includegraphics[width=1\textwidth]{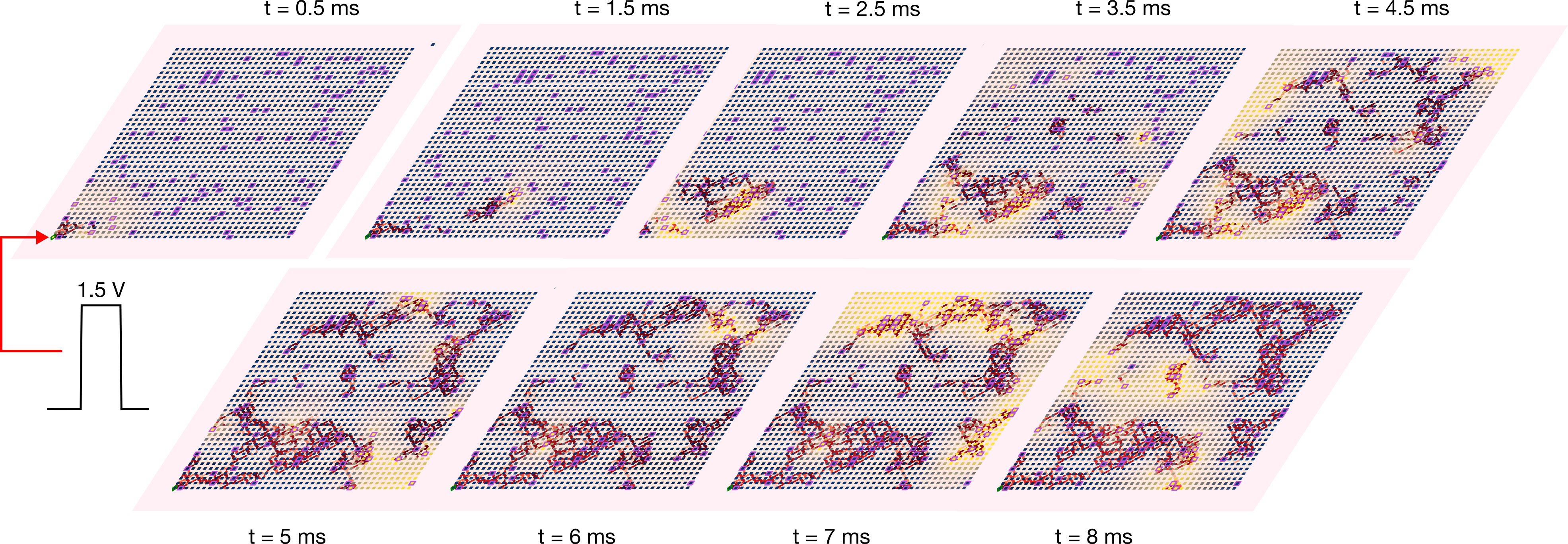}
    \caption{\textbf{Signal propagation through the memristive plexus across space and time}. 
    A voltage pulse of magnitude \qty{1.5}{\volt} and width of \qty{1}{\milli\second} is applied through an input electrode at the bottom left corner of the plexus indicated by the red arrow.
    The signal propagates through the planar system of memristively-coupled \gls{cmos} neurons (depicted by the purple squares). 
    The activation pulse is sufficient to initiate the neurons' self-sustained activity.
    Note how the conductance distribution (red edges) is shaped by the activity of the network, producing clusters in regions with a higher density of \gls{cmos} neurons.
    The size of the simulated system is \qty{1025}{\micro\meter} $\times$ \qty{1025}{\micro\meter}.}
    \label{fig:activity_propagation}
\end{figure*}

The numerical simulation of the Fused-MemBrain hardware is implemented through the
\gls{mvna}~\cite{Chung_Wen_Ho_1975} formalism used in SPICE-type simulators for electronic circuits~\cite{ReviewSpice}. 
\gls{mvna} solves for the currents flowing into the sources and the voltages over each node in the network describing the plexus. 
At each time step, the system is simulated according to the pseudo-code described in~\Cref{sec:pseudocode} where neurons and the plexus interact dynamically.
The implementation of the memristive plexus is based on~\cite{Cipollini2023}, in which the simulator from \cite{montano2022grid} was adapted to take advantage of sparse matrices thanks to the planar nature of the system, therefore reducing simulation time. 
The simulation of the interface between \gls{cmos} neurons and the self-assembled memristive network is an original contribution of this work. 

\subsection{Coarse-graining the plexus}
\label{sec:coarsegraining}

To model the plexus, we aggregate the conductive properties of the self-assembled memristive material deposited on the back-end-of-line.
We tile the plexus in equivalent square regions of $\simeq$ \qty{25}{\micro \meter} linear size equal to the expected unit size of a single \gls{cmos} neuron~\cite{greatorex2024texel}, which becomes our unit scale in space.
In network terminology, a node is defined as the representative of each region of the plexus and is coupled to the adjacent nodes by an edge. 
The topological properties are held simple under the assumption of grid-graph structural connectivity with non-overlapping random diagonal edges to preserve the device's planar dimensionality.
In other words, the conductive properties of the material between two coarse-grained region centers are aggregated and a memristor model is associated with each dynamical edge in the network governing the conductivity level encoded in its weight. 
A similar scheme was originally proposed in Ref.~\cite{montano2022grid,Milano_2021_reservoir} to model conductive networks of nanowires. 

Note that this coarse-graining approach allows the higher-order interactions between \gls{cmos} neurons that would otherwise be lost if we modeled the system as a pairwise interaction network.
Moreover, space and time retain a role in signal propagation delays. 
Delays depend on the dynamical model and the time constants of the memristor model, i.e. the edge connecting two neighboring regions in our model, as well as the space the signal needs to traverse.

\subsection{Memristor model}
\label{sec:memristive_model}

The voltage-driven dynamic of a single memristive edge connecting two coarse-grained regions of the plexus is captured by the model originally proposed for nanowire networks~\cite{MirandaModel,Milano_2021_reservoir,montano2022grid} exhibiting \gls{stp} effects including potentiation, depression and relaxation, paired-pulse-facilitation~\cite{Zucker2002}, as well as heterosynaptic plasticity.
The memristor model is based on the following potentiation-depression rate balance equation:
\begin{equation}
    \label{eq:norm_conductance}
    \frac{dg}{dt} = (1-g)k_p(V) - g k_d(V)
\end{equation}
where $0\leq g \leq1$ is the normalized conductance and $k_p$, $k_d$ are the potentiation and depression rate coefficients. 
For simplicity, they are assumed to depend exponentially only on the absolute voltage difference between two neighboring regions and account for the physical ionic diffusion~\cite{MirandaModel,Wang_2016,Rodriguez-Fernandez,Menzel_2013}:

\begin{equation}
    \label{eq:rate_coeff}
    k_{p,d}(V) = k_{p0, d0}e^{\pm\eta_{p,d}|V|}
\end{equation}
where $+$ and $-$ are associated with $k_p$ and $k_d$ respectively. $k_{p0, d0}>0$ are fitting constants and $\eta_{p,d}>0$ are transition rates.

The current flowing through each edge is assumed to follow Ohm's law for electrical transport:
\begin{equation}
I(t) = [ g(t) G_{\text{max}} + (1-g(t))G_{\text{min}}]V(t)
\label{eq:mem_ohm_law}
\end{equation}
where $G_{\text{min}}$ and $G_{\text{max}}$ are the minimum and maximum values of the conductance, respectively. 
These conductance values need to be empirically identified according to the physical memristive material used for the actual physical implementation of the Fused-MemBrain hardware. 
To develop the empirical realization of the hardware, a more useful and practical quantity would be the conductivity, i.e. the conductance per coarse-grained unit length. 
Specifically, it would be convenient to use an (edge) conductivity $\sigma_* = G_*/l$ over the fundamental length, $l = 25~\mu m$, defined in the coarse-graining by the \gls{cmos} neuron linear size in \Cref{sec:memristive_model}.

The advantage of the memristor model of~\Cref{eq:norm_conductance,eq:rate_coeff} is the low computational cost due to the analytical solution available for discrete time steps $\Delta t>0$:
\begin{equation}
    g(t+\Delta t) =  \Tilde{g} ( 1 - e^{-\theta\Delta t}) + g(t)e^{-\theta\Delta t}
    \label{eq:iterative_norm_cond}
\end{equation}

where $\Tilde{g}(V) = \frac{k_p}{k_p + k_d}$ is the dynamic attractor of the memristive network~\cite{caravelli2023mean,milano2024self}. 
In other words, $\Tilde{g}$ is the final state of the memristive plexus for a given applied voltage. 
Note that such a final state is independent of the initial conductance state.
The exponent $\theta=k_p + k_d$ modulates the speed of reversion and it also depends only on the voltage.
The model has 6 parameters summarized in~\Cref{tab:memristor_parameters} plus an extra parameter that sets the edge pristine conductance.

As a final remark, for edges that do not show memristive properties but only linear electrical conduction, \Cref{eq:mem_ohm_law,eq:iterative_norm_cond} reduce to $g \simeq 0$ and $I(t) \simeq G_{\text{min}}V(t)$, respectively. 
The possibility of setting a uniform distribution of Ohmic edges is included in the simulator settings.
Moreover, our simulator design's modularity and extensibility allow for the easy substitution of distinct memristor models within our simulation scheme, important for comparing and prototyping different self-assembled materials.

\begin{table*}[tb]
\small 
\renewcommand{\arraystretch}{0.8}
\centering
\begin{tabular}{@{} c c c c c l @{}}
\toprule
\textbf{Parameter} & \textbf{Description} & \textbf{Units} & \textbf{Value} \\ \midrule
$k_{p0}$ & Potentiation fitting constant & s$^{-1}$ & \qty{2.56}{\micro\second^{-1}} ($\times 10^{-6}$ s$^{-1}$) \\ 
$k_{d0}$ & Depression fitting constant & s$^{-1}$ & \qty{64.90}{s^{-1}} \\ 
$\eta_p$ & Potentiation transition rate & V$^{-1}$ & \qty{34.90}{V^{-1}} \\ 
$\eta_d$ & Depression transition rate & V$^{-1}$ & \qty{5.59}{V^{-1}} \\ 
$G_{\text{max}}$ & Maximum conductance & S & \qty{200}{pS} (\qty{200}{\times 10^{-12}} S) \\ 
$G_{\text{min}}$ & Minimum conductance & S & \qty{1}{pS} ($\qty{1}{\times 10^{-12}}$ S) \\ \bottomrule
\end{tabular}
\caption{\textbf{Parameters of the memristor model.} $k_{p0},~k_{d0},~\eta_p,~\eta_d$ values are from Ref.~\cite{Milano_2021_reservoir}. 
In Ref.~\cite{Milano_2021_reservoir}, $G_{\text{min}}=1.01$ mS, $G_{\text{max}}=2.72$ mS. 
Here they are adapted to obtain suitable values $I_\text{ext}\simeq$ \qty{10}{\pico \ampere} currents compatible with the \gls{cmos} neurons according to the relation $I_\text{ext}\sim (G_{max}-G_{min})\cdot(A^{(p)}-A^{(n)})$. 
Note that in Ref.~\cite{Milano_2021_reservoir}, $G_{\text{min}}$, $G_{\text{max}}$ are effective values measured for a specific configuration of electrodes.}
\label{tab:memristor_parameters}
\end{table*}

\subsection{Spiking neuron model}

We model and approximate the deterministic \gls{cmos} neuron dynamics with the {lif}LIF neuron model proposed by Lapicque~\cite{Lapicque1907}:

\begin{equation}
    \frac{dV_{m}}{dt} = - \frac{V_{m}}{\tau_{m}} + \frac{I_{\text{ext}}}{C_{m}}
    \label{eq:lif}
\end{equation}

$V_{m}$ is the membrane potential, $\tau_{m}$ is the membrane time constant, $C_{m}$ is the membrane capacitance, and the $I_\text{ext}$ is the external current inflow from the memristive plexus to the \gls{cmos} neuron input electrode.
We approximate \Cref{eq:lif} with the discrete update rule with exponential decay:

\begin{equation}
    V_{m}(t+\Delta t) = V_{m}(t)e^{-\Delta t / \tau_{m}} + \frac{I_\text{ext}(t)}{C_{m}}\Delta t
    \label{eq:discrete_lif}
\end{equation}

When the neuron's membrane potential reaches the spike threshold, $V^{\text{th}}$, it spikes and applies a voltage pulse to the plexus of the type illustrated in panel (a) of \Cref{fig:neuron_conf}.
This pulse is parametrized by the positive and negative pulse widths, $t^{(p)},~t^{(n)}$, and the positive and negative amplitudes, $A^{(p)},~A^{(n)}$.
Immediately after the spike, the membrane potential is reset to $V_m=0$ and the neuron enters its refractory period for a time equal to $t^{(n)}$.
\Cref{tab:lif_parameters} summarizes the parameters of the \gls{cmos} \gls{lif} model used in our simulations.
More complex neuron models could, of course, also be implemented; however, the \gls{lif} neuron model is sufficient as a framework for future \gls{cmos} implementations and investigations into the behavior of this type of hardware.

\begin{table*}[tb]
\small 
\renewcommand{\arraystretch}{0.8}
\centering
\begin{tabular}{ c c c c c c}
\toprule
\textbf{Parameter} & \textbf{Description} & \textbf{Units} & \makecell{\textbf{Typical} \\ \textbf{CMOS value}} & \textbf{Simulation value} & \textbf{Tunable} \\ \midrule
$\tau_m$ & Membrane time constant & s & 10 - 100 ms & 1 ms & \cmark \\ 
$I_\text{ext}$ & External current inflow & A & [10 pA, 1 nA] & $\sim 1 \times 10^{-10}$ A & NA \\ 
$V^{th}$ & Voltage threshold for neuron spike & V & [0.5, 0.9] V & 0.5 V & \cmark  \\ 
$t^{(p)}$ & Positive pulse width (spike) & s & - & 0.50 ms ($5 \times 10^{-4}$ s) & \cmark \\ 
$t^{(n)}$ & Negative pulse width (refractory) & s & - & 0.3 ms ($3 \times 10^{-4}$ s) & \cmark \\ 
$A^{(p)}$ & Positive pulse amplitude (spike) & V & [0.5, 2] V & 1.2 V & \cmark \\
$A^{(n)}$ & Negative pulse amplitude (refractory) & V & [-0.5, -0.1] V & -0.1 V & \cmark \\ 
$V_m$ & Membrane potential & V & [0, $V^{\text{th}}$] V & [0, $V^{th}$] V & NA \\ 
$C_m/\Delta t$ & Membrane capacitance & F/s & - & 3.5 $\times 10^{-20}$ F/s & \xmark \\ 
$\Delta t$ & Discrete time step & s & - & 0.1 ms (0.0001 s) & NA \\ \bottomrule
\end{tabular}
\caption{\textbf{Parameters of the \gls{cmos} \gls{lif} neuron model.} The last column indicates the tunable parameters, which thus provide a mean of control over the dynamical behavior of the overall system, and their possible exploitation in learning.}
\label{tab:lif_parameters}
\end{table*}

\subsection{Simulator algorithm}
\label{sec:pseudocode}

The following algorithm defines how the Fused-MemBrain simulator is both initialized and executed:

\begin{enumerate}
    \item Initialize the plexus with the conductance distribution over the edges.
    \item Set the electrode's location. Place input electrodes and neurons.
    \item Provide the input signal that will drive the network out of equilibrium.
\end{enumerate}

At each time step:

\begin{enumerate}
    \item Identify the neurons that have reached their spiking threshold and prime the voltage pulse that will be applied to the plexus at the next time step.
    \item Apply the voltages computed before and those of the input electrodes and solve the voltage distribution and the current flowing over the plexus.
    \item Update the edge conductances according to the memristor model and the resolved voltage distribution.
    \item Update the membrane potential of the \gls{cmos} neurons based on the in-flowing currents or reset it to the baseline if the neuron produced a spike.
\end{enumerate}
   
\subsection{Example dynamical regime}

\begin{figure*}[hbt]
    \centering
    \includegraphics[width=0.9\linewidth]{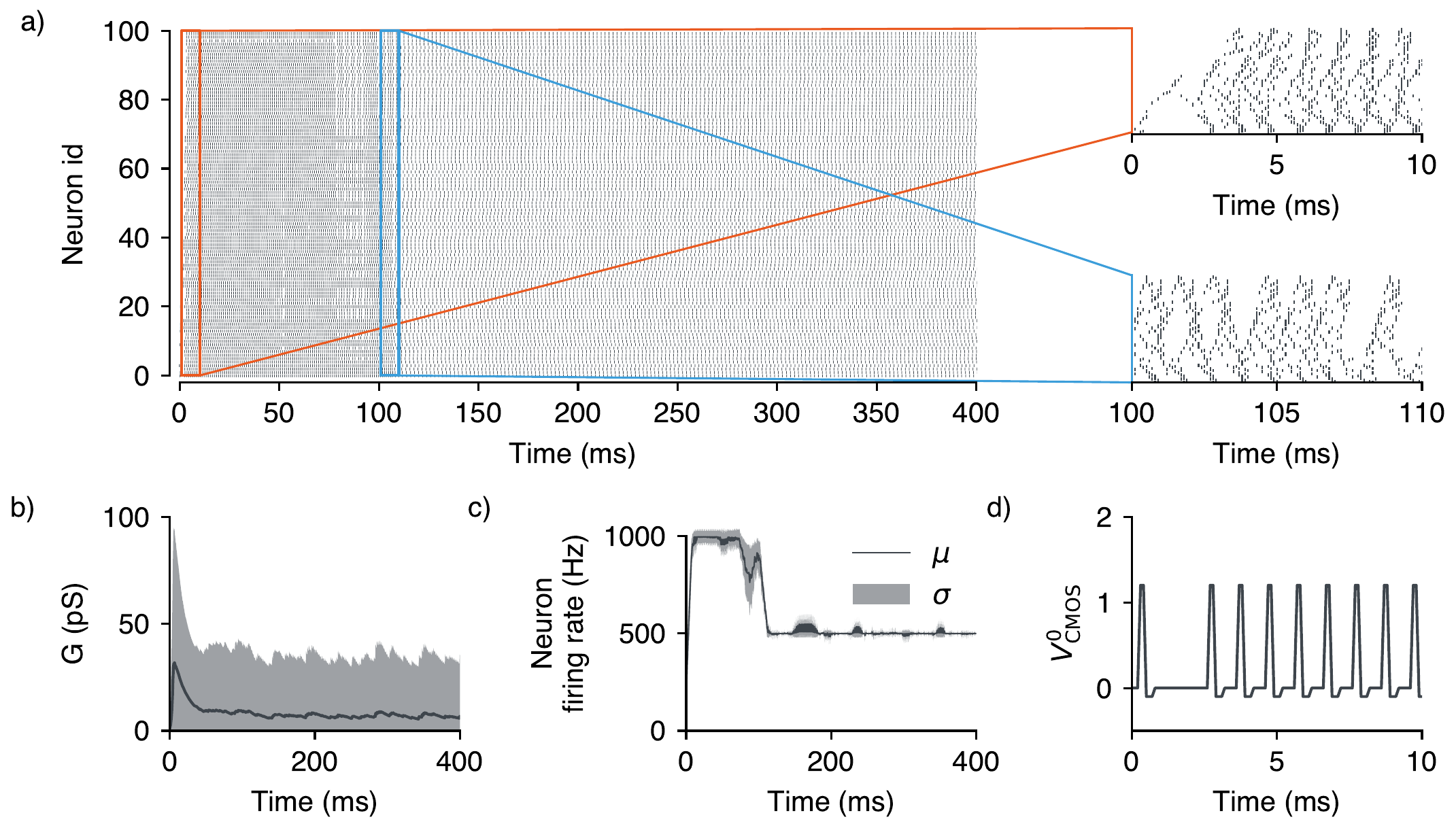}
    \caption{\textbf{Example network activity}. 
    \textbf{a)} The firing activity of the \gls{cmos} neurons. 
    The top inset depicts the initial propagation of the activity, while the bottom inset depicts the activity at a later time.
    \textbf{b)} The self-sustained neural activity raises the average conductivity from the initial pristine value of the plexus. 
    \textbf{c)} The average firing rate of the neurons in the network.
    \textbf{d)} The voltage that neuron 0 applies to the plexus. 
    The neuron begins firing early as it is located close to the origin of the activation signal (bottom left corner of the plexus).}
    \label{fig:activity}
\end{figure*}

In \Cref{fig:activity}, we stimulate the network with a \qty{1.5}{\volt} pulse of \qty{1}{\micro \second} injected at the bottom left corner indicated by the red arrow in \Cref{fig:activity_propagation}. 
The neuron's spiking activity is illustrated in \Cref{fig:activity}a. 
Note the abrupt shift from an initial dense firing activity to one much sparser.  
The top inset on the right of \Cref{fig:activity}a shows the initial firing activity propagating as a wavefront from the nodes closer to the input location which then spreads to the rest of the network through time and space. 
The neuron's firing activity is sustaining and is drawn into the basin of a dynamical attractor state, illustrated in the bottom inset of panel \Cref{fig:activity}a.
Modules of recurrent high-conductive pathways form in the memristive plexus and stabilize after a transient. 
This is evidenced by the direct observation of insets of \Cref{fig:activity_propagation} along with \Cref{fig:activity}b. 
The latter shows the average conductance over the memristive edges in time depicting the interplay between the memristive plexus and the firing rates of the \gls{cmos} neurons. 
\Cref{fig:activity}b shows the average spiking activity of the network starting at high frequency to then relaxing to a lower firing rate.
In \Cref{fig:activity}d, the voltage signal applied to the plexus by the 0-th \gls{cmos} neuron is depicted.

This example demonstrates one possible dynamical regime of the simulator, achieved by tuning neuron parameters and their spatial arrangement. 
A potential use case for this regime could be pattern storage, represented by distinct firing activities across the network, bearing a strong resemblance to attractor networks~\cite{Amit1992,Hopfield1982,Poucet2005}.

\section{Discussion}

Typically, synapses in neuromorphic hardware systems are made of \gls{cmos} elements as much as the neurons. 
Consequently, their design and fabrication require considerable engineering effort because of their bottom-up design.
Moreover, the realization of synapses is usually responsible for high power consumption and significant area overhead~\cite{Qiao2015}. 
In most networks that neuromorphic hardware aims to emulate, the number of synapses scales quadratically with the number of neurons~\cite{Cotteret2024}, making synapses the primary contributor to the cost of fabrication.
Within this work, we envision a potential solution to this problem by endowing neuromorphic hardware with the power of self-assembled memristive materials~\cite{Vahl_2024} that act as a connectivity matrix. 

Understanding and controlling the type of dynamics this system can express is crucial for designing neuromorphic hardware capable of solving tasks.
To this end, we developed a compact open-source simulator (see Code availability) that can be easily used to identify the working conditions in terms of neuron parameters, electrode locations, and compact device models to finally investigate the compatibility of self-assembled materials in combination with \gls{cmos} neuron implementations. 

\subsection{Physical embedding and higher-order interactions}

Neurons communicate with the plexus using digital bipolar pulses as shown in panel \Cref{fig:neuron_conf}a. 
The memristive material (plexus) converts these pulses into currents flowing around the self-assembled network which are sourced into the electrodes of the \gls{cmos} neurons. 
The plexus mediates the interactions between neurons, which may not necessarily be pairwise~\cite{dalgaty2024mosaic}.
In this system, conductive traces are confined in the 2D embedding space shared between the \gls{cmos} neurons, potentially leading to higher order correlations in the activity compared to a network constrained to pair-wise connections.
Moreover, this influences the emergence of possible topological traits, such as modules and clusters, self-organized by the activity above the \gls{cmos}.
Networks, where connections between nodes incur a cost for spanning physical distances, tend to exhibit higher clustering and fewer high-degree nodes. 
This pattern arises as a natural consequence of their Euclidean embedding~\cite{barthelemy2022spatial,Barthlemy2003,Kaiser2004}.
Similarly, the physical structure of the brain~\cite{zhang2024geometric} constrains its network topology, resulting in a clustered organization with few long-range connections, yet it displays remarkable computation power.
Whether this computational efficiency is a direct outcome of these constraints remains an open and interesting question. 
With this in mind, it makes sense to explore hardware that embraces the physical constraints instead of attempting to circumvent them with costly and challenging connectivity infrastructures.
Such hardware could be more economically feasible and cheaper to engineer while still benefiting from high-performance information processing. 

\subsection{Pathways for system use}

An open question raised in this paper is how to effectively utilize such a system.
We explore several potential approaches and discuss a selection of them below.

A typical approach would be to use the reservoir computing paradigm, also known as echo-state networks~\cite{jaeger2001echo} or \glspl{lsm} (when units are spiking)~\cite{Maass2007}.
The widely adopted interest in these networks is due to their simplicity and ability to tame dynamical systems into performing useful computation. 
Only the weights connecting the recursive and output layers need to be learned, usually employing an \gls{svm} or an \gls{mlp}.
To effectively use the Fused-MemBrain hardware within this context, the temporal behavior of the system needs to be compatible with that of the input data.
Therefore, the firing activity of the hardware neurons should be tuned such that the overall system exhibits a decaying behavior after being driven out of equilibrium with the input data.
Similar to reservoir computing, \textcite{Henseler1992} used the differential equations describing the dynamics of a two-dimensional sheet (called ``membrain'') to generate moving waves under an input signal perturbation. 
The ``membrain'' dynamics were then used as the input for a linear classifier to solve a recognition task.
This closely resembles the dynamics observed in our proposed system, as depicted in \Cref{fig:activity_propagation}.

In comparison to the reservoir scheme, where the activity is in the decaying regime, another possibility is to use the system in the \textit{self-sustained} regime as shown in \Cref{fig:activity}.
In this operation mode, learning could be achieved by ``structuring'' the waves to be pattern-dependent leveraging the Hebbian-like plasticity expressed by the memristive plexus.

In recent work~\cite{kagan2022vitro,isomura2018vitro,isomura2015cultured} \textit{in vitro} biological networks coupled with a multi-electrode array were able to demonstrate ``learning'' in the context of a closed-loop system.
Their experimental setup bears resemblance to the system proposed in this work, in which a high-density network (the plexus) is interfaced with a high-density electrode array (\gls{cmos} neurons).
How to exhibit a similar demonstration with the Fused-MemBrain hardware depends on the properties of the memristive material used in the physical realization.
For instance, when the plexus shows memristive volatile behavior the main control we have over the system dynamics are the neuron parameters, input electrodes, and their locations on the plexus.
The volatility of the plexus does not allow for the retention of connectivity structures imperative for a given task, any change in conductivity, and therefore network connectivity, will be lost as the system has an inertia to remain in its pristine state. 
However, such volatile dynamics remain useful to route signals between neurons. 
The neurons can be tuned by changing their parameters to impose desired wave patterns within the system.
Learning would then be achieved by the dynamics of the neurons existing on multiple time scales as well as their spatial inhomogeneity.
As an example, this could be implemented by ensuring neuron activity aligns with the timescales exhibited by the plexus, maintaining an elevated connectivity state and preventing it from completely relaxing. 

Alternatively, should the plexus show non-volatile behavior we can design the inhomogeneity of its conductive properties externally, sending the desired signals to write or erase any high-conductivity trace formation in the plexus. 
This is similar to the approach adopted in Ref.~\cite{hughes2019wave} where a physical wave system was trained to learn complex features in temporal data. 
They demonstrated the classification of audio signals using waveform scattering and propagation through an inhomogeneous medium.
For the Fused-MemBrain hardware to exploit this scheme it would be necessary that the neuron activity is tuned such that it does not override the engineered inhomogeneity. 

\section{Conclusions}

In this paper, we proposed a novel type of neuromorphic hardware that marries the high-density connectivity of self-assembled memristive systems with the versatility and maturity of \gls{cmos} technology.
The architectural simplicity of our proposal relies on the memristive self-assembled material replacing the synaptic connections between \gls{cmos} neurons in the hardware.
Synaptic connections are accountable for much of the silicon real estate and therefore the economic cost of neuromorphic hardware.
Replacing synapses with a memristive plexus, i.e. a planar sheet of self-assembled memristive elements, with which \gls{cmos} neurons are interfaced, could offer a low-cost solution to signal routing.
The proposed hardware enables the exploration of emerging self-assembled materials interfacing with a high density of electrodes
Simultaneously it offers a platform to investigate spiking neural network topologies that support higher-order interactions, akin to those observed in biological systems.
To pave the way, we developed an open-access simulator to evaluate the compatibility of self-assembled materials with the proposed hardware description, which we believe could ultimately realize information processing directly within the physical domain.

\section*{Code availability}
The simulator code can be accessed on GitHub: \url{https://github.com/CipolliniDavide/FusedMemBrain.git}.

\section*{Acknowledgments}
The authors are grateful to Karolina Tran for her insightful discussions on the potential of self-assembled materials, which greatly enriched this work. 
The authors acknowledge the International Workshop ''Bio-inspired Information Pathways`` by the CRC 1461 in collaboration with partners from CogniGron, as the event that inspired this research.
The authors gratefully acknowledge the financial support from the Groningen Cognitive Systems and Materials Center (CogniGron).
Partially funded by the Deutsche Forschungsgemeinschaft (DFG, German Research Foundation): Project MemTDE Project number 441959088 as part of the DFG priority program SPP 2262 MemrisTec Project number 422738993.
D.C., L.S., and E.C. gratefully acknowledge the EUs Horizon 2020, from the MSCA-ITN-2019 Innovative Training Networks program ”Materials for Neuromorphic Circuits” (MANIC) under the grant agreement No. 861153 for financial support.

\printbibliography

@article{waldrop2016chips,
  title={The chips are down for Moore’s law},
  author={Waldrop, M Mitchell},
  journal={Nature News},
  volume={530},
  number={7589},
  pages={144},
  year={2016},
doi={10.1038/530144a}
}

@article{patterson_carbon_2021,
  title={Carbon emissions and large neural network training},
  author={Patterson, David and Gonzalez, Joseph and Le, Quoc and Liang, Chen and Munguia, Lluis-Miquel and Rothchild, Daniel and So, David and Texier, Maud and Dean, Jeff},
  journal={arXiv preprint arXiv:2104.10350},
  year={2021},
DOI={10.48550/arXiv.2104.10350}
}

@article{Markovi2020,
  title = {Physics for neuromorphic computing},
  volume = {2},
  ISSN = {2522-5820},
  url = {http://dx.doi.org/10.1038/s42254-020-0208-2},
  DOI = {10.1038/s42254-020-0208-2},
  number = {9},
  journal = {Nature Reviews Physics},
  publisher = {Springer Science and Business Media LLC},
  author = {Marković,  Danijela and Mizrahi,  Alice and Querlioz,  Damien and Grollier,  Julie},
  year = {2020},
  month = jul,
  pages = {499–510}
}

@inproceedings{McKee2004,
  doi = {10.1145/977091.977115},
  url = {https://doi.org/10.1145/977091.977115},
  year = {2004},
  month = apr,
  publisher = {{ACM}},
  author = {Sally A. McKee},
  title = {Reflections on the memory wall},
  booktitle = {Proceedings of the 1st conference on Computing frontiers}
}

@article{Mead2020,
  doi = {10.1038/s41928-020-0448-2},
  url = {https://doi.org/10.1038/s41928-020-0448-2},
  year = {2020},
  publisher = {Springer Science and Business Media {LLC}},
  volume = {3},
  number = {7},
  pages = {434--435},
  author = {Carver Mead},
  title = {How we created neuromorphic engineering},
  journal = {Nature Electronics}
}

@article{indiveri2011neuromorphic,
  title={Neuromorphic silicon neuron circuits},
  author={Indiveri, Giacomo and Linares-Barranco, Bernab{\'e} and Hamilton, Tara Julia and Schaik, Andr{\'e} van and Etienne-Cummings, Ralph and Delbruck, Tobi and Liu, Shih-Chii and Dudek, Piotr and H{\"a}fliger, Philipp and Renaud, Sylvie and others},
  journal={Frontiers in neuroscience},
  volume={5},
  pages={73},
  year={2011},
  publisher={Frontiers Research Foundation},
  doi={10.3389/fnins.2011.00073}
}

@article{Burkhardt2023,
  doi = {10.1126/science.ade5645},
  url = {https://doi.org/10.1126/science.ade5645},
  year = {2023},
  month = apr,
  publisher = {American Association for the Advancement of Science ({AAAS})},
  volume = {380},
  number = {6642},
  pages = {293--297},
  author = {Pawel Burkhardt and Jeffrey Colgren and Astrid Medhus and Leonid Digel and Benjamin Naumann and Joan J. Soto-Angel and Eva-Lena Nordmann and Maria Y. Sachkova and Maike Kittelmann},
  title = {Syncytial nerve net in a ctenophore adds insights on the evolution of nervous systems},
  journal = {Science}
}

@article{Qiao2015,
  title = {A reconfigurable on-line learning spiking neuromorphic processor comprising 256 neurons and 128K synapses},
  volume = {9},
  ISSN = {1662-453X},
  doi = {10.3389/fnins.2015.00141},
  journal = {Frontiers in Neuroscience},
  publisher = {Frontiers Media SA},
  author = {Qiao,  Ning and Mostafa,  Hesham and Corradi,  Federico and Osswald,  Marc and Stefanini,  Fabio and Sumislawska,  Dora and Indiveri,  Giacomo},
  year = {2015},
  month = apr 
}

@article{Chicca2014,
  author={Chicca, Elisabetta and Stefanini, Fabio and Bartolozzi, Chiara and Indiveri, Giacomo},
  journal={Proceedings of the IEEE}, 
  title={Neuromorphic Electronic Circuits for Building Autonomous Cognitive Systems}, 
  year={2014},
  volume={102},
  number={9},
  pages={1367-1388},
  doi={10.1109/JPROC.2014.2313954}}

@book{Golgi1885,
  title     = {Sulla fina anatomia degli organi centrali del sistema nervosa},
  author    = {Golgi, Camillo},
  year      = {1885},
  publisher = {Revista sperimentale di Freniatria. Reprinted in: Golgi, C. Opera Omnia. Milano, Hoepli: 1903},
  address   = {Milano}
}

@article{Glickstein2006,
  doi = {10.1016/j.cub.2006.02.053},
  url = {https://doi.org/10.1016/j.cub.2006.02.053},
  year = {2006},
  month = mar,
  publisher = {Elsevier {BV}},
  volume = {16},
  number = {5},
  pages = {R147--R151},
  author = {Mitch Glickstein},
  title = {Golgi and Cajal: The neuron doctrine and the 100th anniversary of the 1906 Nobel Prize},
  journal = {Current Biology}
}

@article{Henseler1992,
  doi = {10.1002/cta.4490200505},
  url = {https://doi.org/10.1002/cta.4490200505},
  year = {1992},
  month = sep,
  publisher = {Wiley},
  volume = {20},
  number = {5},
  pages = {483--496},
  author = {J. Henseler and P. J. Braspenning},
  title = {Membrain: A cellular neural network model based on a vibrating 
           membrane},
  journal = {International Journal of Circuit Theory and Applications}
}

@article{Palade54,
  author  = {G. E. Palade and S. Palay},
  title   = {Electron Microscope Observations of Intraneuronal and Neuromuscular Synapses},
  journal = {Anat. Rec.},
  year    = {1954},
  volume  = {118},
  pages   = {335–336}
}

@article{Gray59,
 author={E. G. Gray},
 title={Axo-somatic and axo-dendritic synapses of the cerebral cortex: an electron microscope study},
 journal={Journal of Anatomy},
 year={1959},
 volume={93(Pt 4)},
 pages={420-33} 
 }

@article{Thoma2023,
  doi = {10.1073/pnas.2221493120},
  url = {https://doi.org/10.1073/pnas.2221493120},
  year = {2023},
  month = apr,
  publisher = {Proceedings of the National Academy of Sciences},
  volume = {120},
  number = {15},
  author = {Vladimiros Thoma and Shuhei Sakai and Koki Nagata and Yuu Ishii and Shinichiro Maruyama and Ayako Abe and Shu Kondo and Masakado Kawata and Shun Hamada and Ryusaku Deguchi and Hiromu Tanimoto},
  title = {On the origin of appetite: {GLWamide} in jellyfish represents an ancestral satiety neuropeptide},
  journal = {Proceedings of the National Academy of Sciences}
}

@INPROCEEDINGS{ReviewSpice,
  author={Pratap, Rajendra and Agarwal, Vineeta and Singh, R K},
  booktitle={2014 International Conference on Power, Control and Embedded Systems (ICPCES)}, 
  title={Review of various available spice simulators}, 
  year={2014},
  volume={},
  number={},
  pages={1-6},
  doi={10.1109/ICPCES.2014.7062809}}

@article{Chung_Wen_Ho_1975,
	doi = {10.1109/tcs.1975.1084079},
	year = 1975,
	publisher = {Institute of Electrical and Electronics Engineers ({IEEE})},
	volume = {22},
	number = {6},
	pages = {504--509},
	author = {Chung-Wen Ho and A. Ruehli and P. Brennan},
	title = {The modified nodal approach to network analysis},
	journal = {{IEEE} Transactions on Circuits and Systems}
}

@article{Cipollini2023,
  doi = {10.1088/2634-4386/acd6b3},
  url = {https://doi.org/10.1088/2634-4386/acd6b3},
  year = {2023},
  month = may,
  publisher = {{IOP} Publishing},
  author = {Davide Cipollini and Lambert Schomaker},
  title = {Conduction and entropy analysis of a mixed memristor-resistor model for neuromorphic networks},
  journal = {Neuromorphic Computing and Engineering}
}

@article{Sillin2013,
  title = {A theoretical and experimental study of neuromorphic atomic switch networks for reservoir computing},
  volume = {24},
  ISSN = {1361-6528},
  url = {http://dx.doi.org/10.1088/0957-4484/24/38/384004},
  DOI = {10.1088/0957-4484/24/38/384004},
  number = {38},
  journal = {Nanotechnology},
  publisher = {IOP Publishing},
  author = {Sillin,  Henry O and Aguilera,  Renato and Shieh,  Hsien-Hang and Avizienis,  Audrius V and Aono,  Masakazu and Stieg,  Adam Z and Gimzewski,  James K},
  year = {2013},
  month = sep,
  pages = {384004}
}

@article{montano2022grid,
  title = {Grid-graph modeling of emergent neuromorphic dynamics and heterosynaptic plasticity in memristive nanonetworks},
  volume = {2},
  ISSN = {2634-4386},
  url = {http://dx.doi.org/10.1088/2634-4386/ac4d86},
  DOI = {10.1088/2634-4386/ac4d86},
  number = {1},
  journal = {Neuromorphic Computing and Engineering},
  publisher = {IOP Publishing},
  author = {Montano,  Kevin and Milano,  Gianluca and Ricciardi,  Carlo},
  year = {2022},
  month = feb,
  pages = {014007}
}

@article{MirandaModel,  
    author={Miranda, Enrique and Milano, Gianluca and Ricciardi, Carlo},  
    journal={IEEE Transactions on Nanotechnology},   
    title={Modeling of Short-Term Synaptic Plasticity Effects in ZnO Nanowire-Based Memristors Using a Potentiation-Depression Rate Balance Equation},   
    year={2020},  
    volume={19},  
    number={},  
    pages={609-612},  
    doi={10.1109/TNANO.2020.3009734}
}

@article{Milano_2021_reservoir,
	doi = {10.1038/s41563-021-01099-9},
	year = 2021,
	publisher = {Springer Science and Business Media {LLC}},
	volume = {21},
	number = {2},
	pages = {195--202},
	author = {Gianluca Milano and Giacomo Pedretti and Kevin Montano and Saverio Ricci and Shahin Hashemkhani and Luca Boarino and Daniele Ielmini and Carlo Ricciardi},
	title = {In materia reservoir computing with a fully memristive architecture based on self-organizing nanowire networks},
	journal = {Nature Materials}
}

@article{rieck2022ferroelastic,
	doi = {10.1002/aisy.202200292},
	year = 2022,
	publisher = {Wiley},
	pages = {2200292},
	author = {Jan L. Rieck and Davide Cipollini and Mart Salverda and Cynthia P. Quinteros and Lambert R. B. Schomaker and Beatriz Noheda},
	title = {Ferroelastic Domain Walls in {BiFeO}$_3$ as Memristive Networks},
	journal = {Advanced Intelligent Systems}
}

@article{Profumo_2023,
	doi = {10.1088/1361-6463/acd704},
	url = {https://doi.org/10.1088%2F1361-6463%2Facd704},
	year = 2023,
	publisher = {{IOP} Publishing},
	volume = {56},
	number = {35},
	pages = {355301},
	author = {Filippo Profumo and Francesca Borghi and Andrea Falqui and Paolo Milani},
	title = {Potentiation and depression behaviour in a two-terminal memristor based on nanostructured bilayer {ZrO}\textsubscript{x}
		               /Au films},
	journal = {Journal of Physics D: Applied Physics}
}

@article{Chen2020,
  title = {Classification with a disordered dopant-atom network in silicon},
  volume = {577},
  ISSN = {1476-4687},
  url = {http://dx.doi.org/10.1038/s41586-019-1901-0},
  DOI = {10.1038/s41586-019-1901-0},
  number = {7790},
  journal = {Nature},
  publisher = {Springer Science and Business Media LLC},
  author = {Chen,  Tao and van Gelder,  Jeroen and van de Ven,  Bram and Amitonov,  Sergey V. and de Wilde,  Bram and Ruiz Euler,  Hans-Christian and Broersma,  Hajo and Bobbert,  Peter A. and Zwanenburg,  Floris A. and van der Wiel,  Wilfred G.},
  year = {2020},
  month = jan,
  pages = {341–345}
}

@article{Mallinson2024,
  title={Experimental Demonstration of Reservoir Computing with Self-Assembled Percolating Networks of Nanoparticles},
  author={Mallinson, Joshua B and Steel, Jamie K and Heywood, Zachary E and Studholme, Sofie J and Bones, Philip J and Brown, Simon A},
  journal={Advanced Materials},
  pages={2402319},
  year={2024},
  publisher={Wiley Online Library}, 
  doi={doi.org/10.1002/adma.202402319}
}

@article{Zhu2023,
  title = {Online dynamical learning and sequence memory with neuromorphic nanowire networks},
  volume = {14},
  ISSN = {2041-1723},
  url = {http://dx.doi.org/10.1038/s41467-023-42470-5},
  DOI = {10.1038/s41467-023-42470-5},
  number = {1},
  journal = {Nature Communications},
  publisher = {Springer Science and Business Media LLC},
  author = {Zhu,  Ruomin and Lilak,  Sam and Loeffler,  Alon and Lizier,  Joseph and Stieg,  Adam and Gimzewski,  James and Kuncic,  Zdenka},
  year = {2023},
  month = nov 
    }

@article{Lapicque1907, 
    author = {Lapicque, Louis},
    title={Recherches quantitatives sur l'excitation électrique des nerfs traitée comme une polarisation},
    journal={J. Physiol. Pathol.},
    year={1907},
    volume={9},
    pages={620–635}
    }

@article{Pallasdies2019,
  doi = {10.7554/elife.50084},
  url = {https://doi.org/10.7554/elife.50084},
  year = {2019},
  month = dec,
  publisher = {{eLife} Sciences Publications,  Ltd},
  volume = {8},
  author = {Fabian Pallasdies and Sven Goedeke and Wilhelm Braun and Raoul-Martin Memmesheimer},
  title = {From single neurons to behavior in the jellyfish Aurelia aurita},
  journal = {{eLife}}
}

@article{Yuan2014,
  doi = {10.4208/aamm.2013.m409},
  url = {https://doi.org/10.4208/aamm.2013.m409},
  year = {2014},
  month = jun,
  publisher = {Global Science Press},
  volume = {6},
  number = {3},
  pages = {307--326},
  author = {Hai-Zhuan Yuan and Shi Shu and Xiao-Dong Niu and Mingjun Li and Yang Hu},
  title = {A Numerical Study of Jet Propulsion of an Oblate Jellyfish Using a Momentum Exchange-Based Immersed Boundary-Lattice {Boltzmann} Method},
  journal = {Advances in Applied Mathematics and Mechanics}
}

@article{remacle1998networks,
  title={Networks of quantum nanodots: The role of disorder in modifying electronic and optical properties},
  author={Remacle, F and Collier, CP and Markovich, G and Heath, JR and Banin, U and Levine, RD},
  journal={The Journal of Physical Chemistry B},
  volume={102},
  number={40},
  pages={7727--7734},
  year={1998},
  publisher={ACS Publications},
  doi={10.1021/jp9813948}
}

@article{liu2021colloidal,
  title={Colloidal quantum dot electronics},
  author={Liu, Mengxia and Yazdani, Nuri and Yarema, Maksym and Jansen, Maximilian and Wood, Vanessa and Sargent, Edward H},
  journal={Nature Electronics},
  volume={4},
  number={8},
  pages={548--558},
  year={2021},
  publisher={Nature Publishing Group UK London},
  doi = {10.1038/s41928-021-00632-7}
}

@article{lynch1977heterosynaptic,
  title={Heterosynaptic depression: a postsynaptic correlate of long-term potentiation},
  author={Lynch, Gary S and Dunwiddie, Thomas and Gribkoff, Valentin},
  journal={Nature},
  volume={266},
  number={5604},
  pages={737--739},
  year={1977},
  publisher={Nature Publishing Group UK London},
  doi={10.1038/266737a0}
}

@article{Wang_2016,
	doi = {10.1038/nmat4756},
	year = 2016,
	publisher = {Springer Science and Business Media {LLC}},
	volume = {16},
	number = {1},
	pages = {101--108},
	author = {Zhongrui Wang and Saumil Joshi and Sergey E. Savel'ev and Hao Jiang and Rivu Midya and Peng Lin and Miao Hu and Ning Ge and John Paul Strachan and Zhiyong Li and Qing Wu and Mark Barnell and Geng-Lin Li and Huolin L. Xin and R. Stanley Williams and Qiangfei Xia and J. Joshua Yang},
	title = {Memristors with diffusive dynamics as synaptic emulators for neuromorphic computing},
	journal = {Nature Materials}
}

@article{Menzel_2013,
	doi = {10.1039/c3cp50738f},
	year = 2013,
	publisher = {Royal Society of Chemistry (RSC)},
	volume = {15},
	number = {18},
	pages = {6945},
	author = {Stephan Menzel and Stefan Tappertzhofen and Rainer Waser and Ilia Valov},
	title = {Switching kinetics of electrochemical metallization memory cells},
	journal = {Physical Chemistry Chemical Physics}
}

@article{Rodriguez-Fernandez,  
    author={Rodriguez-Fernandez, A. and Cagli, C. and Suñe, J. and Miranda, E.},   journal={IEEE Electron Device Letters},   
    title={Switching Voltage and Time Statistics of Filamentary Conductive Paths in {HfO}$_2$-Based {ReRAM} Devices},   
    year={2018},  
    volume={39},  
    number={5},  
    pages={656-659},  
    doi={10.1109/LED.2018.2822047}
}

@article{milano2024self,
  title={Self-organizing neuromorphic nanowire networks are stochastic dynamical systems},
  author={Milano, Gianluca and Michieletti, Fabio and Ricciardi, Carlo and Miranda, Enrique},
  year={2024},
  journal={Research Square},
  doi={https://doi.org/10.21203/rs.3.rs-4102090/v1}
}

@article{caravelli2023mean,
  title = {Mean Field Theory of Self‐Organizing Memristive Connectomes},
  volume = {535},
  ISSN = {1521-3889},
  url = {http://dx.doi.org/10.1002/andp.202300090},
  DOI = {10.1002/andp.202300090},
  number = {8},
  journal = {Annalen der Physik},
  publisher = {Wiley},
  author = {Caravelli,  Francesco and Milano,  Gianluca and Ricciardi,  Carlo and Kuncic,  Zdenka},
  year = {2023},
  month = jun 
}

@article{Vahl_2024,
doi = {10.1088/1361-6463/ad7a82},
url = {https://dx.doi.org/10.1088/1361-6463/ad7a82},
year = {2024},
month = sep,
publisher = {IOP Publishing},
volume = {57},
number = {50},
pages = {503001},
author = {Alexander Vahl and Gianluca Milano and Zdenka Kuncic and Simon A Brown and Paolo Milani},
title = {Brain-inspired computing with self-assembled networks of nano-objects},
journal = {Journal of Physics D: Applied Physics},
}

@article{jaeger2001echo,
  title={The “echo state” approach to analysing and training recurrent neural networks-with an erratum note},
  author={Jaeger, Herbert},
  journal={Bonn, Germany: German National Research Center for Information Technology GMD Technical Report},
  volume={148},
  number={34},
  pages={13},
  year={2001},
  publisher={Bonn},
  url={https://api.semanticscholar.org/CorpusID:15467150}
}

@article{kagan2022vitro,
  title={In vitro neurons learn and exhibit sentience when embodied in a simulated game-world},
  author={Kagan, Brett J and Kitchen, Andy C and Tran, Nhi T and Habibollahi, Forough and Khajehnejad, Moein and Parker, Bradyn J and Bhat, Anjali and Rollo, Ben and Razi, Adeel and Friston, Karl J},
  journal={Neuron},
  volume={110},
  number={23},
  pages={3952--3969},
  year={2022},
  publisher={Elsevier},
  doi={10.1016/j.neuron.2022.09.001}
}

@article{hughes2019wave,
  title = {Wave physics as an analog recurrent neural network},
  volume = {5},
  ISSN = {2375-2548},
  url = {http://dx.doi.org/10.1126/sciadv.aay6946},
  DOI = {10.1126/sciadv.aay6946},
  number = {12},
  journal = {Science Advances},
  publisher = {American Association for the Advancement of Science (AAAS)},
  author = {Hughes,  Tyler W. and Williamson,  Ian A. D. and Minkov,  Momchil and Fan,  Shanhui},
  year = {2019},
  month = dec 
}

@article{dalgaty2024mosaic,
  title={Mosaic: in-memory computing and routing for small-world spike-based neuromorphic systems},
  author={Dalgaty, Thomas and Moro, Filippo and Demira{\u{g}}, Yi{\u{g}}it and De Pra, Alessio and Indiveri, Giacomo and Vianello, Elisa and Payvand, Melika},
  journal={Nature Communications},
  volume={15},
  number={1},
  pages={142},
  year={2024},
  publisher={Nature Publishing Group UK London},
  doi={10.1038/s41467-023-44365-x}
}

@article{bullmore2009complex,
  title={Complex brain networks: graph theoretical analysis of structural and functional systems},
  author={Bullmore, Ed and Sporns, Olaf},
  journal={Nature reviews neuroscience},
  volume={10},
  number={3},
  pages={186--198},
  year={2009},
  publisher={Nature Publishing Group UK London},
  doi={10.1038/nrn2575}
}

@article{barabasi1999emergence,
  title={Emergence of scaling in random networks},
  author={Barab{\'a}si, Albert-L{\'a}szl{\'o} and Albert, R{\'e}ka},
  journal={science},
  volume={286},
  number={5439},
  pages={509--512},
  year={1999},
  publisher={American Association for the Advancement of Science},
  doi={10.1126/science.286.5439.509}
}

@article{kramer2008emergent,
  title={Emergent network topology at seizure onset in humans},
  author={Kramer, Mark A and Kolaczyk, Eric D and Kirsch, Heidi E},
  journal={Epilepsy research},
  volume={79},
  number={2-3},
  pages={173--186},
  year={2008},
  publisher={Elsevier},
  doi={10.1016/j.eplepsyres.2008.02.002}
}

@article{wang2009altered,
  title={Altered small-world brain functional networks in children with attention-deficit/hyperactivity disorder},
  author={Wang, Liang and Zhu, Chaozhe and He, Yong and Zang, Yufeng and Cao, QingJiu and Zhang, Han and Zhong, Qiuhai and Wang, Yufeng},
  journal={Human brain mapping},
  volume={30},
  number={2},
  pages={638--649},
  year={2009},
  publisher={Wiley Online Library},
  doi={doi.org/10.1002/hbm.20530}
}

@article{liu2008disrupted,
  title={Disrupted small-world networks in schizophrenia},
  author={Liu, Yong and Liang, Meng and Zhou, Yuan and He, Yong and Hao, Yihui and Song, Ming and Yu, Chunshui and Liu, Haihong and Liu, Zhening and Jiang, Tianzi},
  journal={Brain},
  volume={131},
  number={4},
  pages={945--961},
  year={2008},
  publisher={Oxford University Press},
  doi={10.1093/brain/awn018}
}

@article{micheloyannis2006small,
  title={Small-world networks and disturbed functional connectivity in schizophrenia},
  author={Micheloyannis, Sifis and Pachou, Ellie and Stam, Cornelis Jan and Breakspear, Michael and Bitsios, Panagiotis and Vourkas, Michael and Erimaki, Sophia and Zervakis, Michael},
  journal={Schizophrenia research},
  volume={87},
  number={1-3},
  pages={60--66},
  year={2006},
  publisher={Elsevier},
  doi={10.1016/j.schres.2006.06.028}
}

@article{barthelemy2011spatial,
  title={Spatial networks},
  author={Barth{\'e}lemy, Marc},
  journal={Physics reports},
  volume={499},
  number={1-3},
  pages={1--101},
  year={2011},
  publisher={Elsevier},
  doi={10.1016/j.physrep.2010.11.002}
}

@article{zhang2024geometric,
  title={Geometric scaling law in real neuronal networks},
  author={Zhang, Xin-Ya and Moore, Jack Murdoch and Ru, Xiaolei and Yan, Gang},
  journal={Physical Review Letters},
  volume={133},
  number={13},
  pages={138401},
  year={2024},
  publisher={APS},
  doi={10.1103/PhysRevLett.133.138401}
}

@article{isomura2015cultured,
  title={Cultured cortical neurons can perform blind source separation according to the free-energy principle},
  author={Isomura, Takuya and Kotani, Kiyoshi and Jimbo, Yasuhiko},
  journal={PLoS computational biology},
  volume={11},
  number={12},
  pages={e1004643},
  year={2015},
  publisher={Public Library of Science San Francisco, CA USA},
  doi={doi.org/10.1371/journal.pcbi.1004643}
}

@article{isomura2018vitro,
  title={In vitro neural networks minimise variational free energy},
  author={Isomura, Takuya and Friston, Karl},
  journal={Scientific reports},
  volume={8},
  number={1},
  pages={16926},
  year={2018},
  publisher={Nature Publishing Group UK London},
  doi={10.1038/s41598-018-35221-w}
}

@article{friston2010free,
  title={The free-energy principle: a unified brain theory?},
  author={Friston, Karl},
  journal={Nature reviews neuroscience},
  volume={11},
  number={2},
  pages={127--138},
  year={2010},
  publisher={Nature publishing group},
  doi={/doi.org/10.1038/nrn2787}
}

@misc{greatorex2024texel,
      title={{TEXEL}: A neuromorphic processor with on-chip learning for beyond-{CMOS} device integration}, 
      author={Hugh Greatorex and Ole Richter and Michele Mastella and Madison Cotteret and Philipp Klein and Maxime Fabre and Arianna Rubino and Willian Soares Girão and Junren Chen and Martin Ziegler and Laura Bégon-Lours and Giacomo Indiveri and Elisabetta Chicca},
      year={2024},
      eprint={2410.15854},
      archivePrefix={arXiv},
}

@article{Zucker2002,
  title = {Short-Term Synaptic Plasticity},
  volume = {64},
  ISSN = {1545-1585},
  url = {http://dx.doi.org/10.1146/annurev.physiol.64.092501.114547},
  DOI = {10.1146/annurev.physiol.64.092501.114547},
  number = {1},
  journal = {Annual Review of Physiology},
  publisher = {Annual Reviews},
  author = {Zucker,  Robert S. and Regehr,  Wade G.},
  year = {2002},
  month = mar,
  pages = {355–405}
}

@article{Poucet2005,
  title = {Attractors in Memory},
  volume = {308},
  ISSN = {1095-9203},
  url = {http://dx.doi.org/10.1126/science.1112555},
  DOI = {10.1126/science.1112555},
  number = {5723},
  journal = {Science},
  publisher = {American Association for the Advancement of Science (AAAS)},
  author = {Poucet,  Bruno and Save,  Etienne},
  year = {2005},
  month = may,
  pages = {799–800}
}

@book{Amit1992,
author = {Amit, Daniel J.},
title = {Modelling Brain Function: The World of Attractor Neural Networks},
year = {1992},
isbn = {0521421241},
publisher = {Cambridge University Press},
address = {USA},
edition = {1st}
}

@article{Hopfield1982,
  title = {Neural networks and physical systems with emergent collective computational abilities.},
  volume = {79},
  ISSN = {1091-6490},
  url = {http://dx.doi.org/10.1073/pnas.79.8.2554},
  DOI = {10.1073/pnas.79.8.2554},
  number = {8},
  journal = {Proceedings of the National Academy of Sciences},
  publisher = {Proceedings of the National Academy of Sciences},
  author = {Hopfield,  J J},
  year = {1982},
  month = apr,
  pages = {2554–2558}
}

@article{Cotteret2024,
  title = {Vector Symbolic Finite State Machines in Attractor Neural Networks},
  volume = {36},
  ISSN = {1530-888X},
  url = {http://dx.doi.org/10.1162/neco_a_01638},
  DOI = {10.1162/neco_a_01638},
  number = {4},
  journal = {Neural Computation},
  publisher = {MIT Press},
  author = {Cotteret,  Madison and Greatorex,  Hugh and Ziegler,  Martin and Chicca,  Elisabetta},
  year = {2024},
  month = mar,
  pages = {549–595}
}

@book{barthelemy2022spatial,
  title = {Spatial Networks: A Complete Introduction: From Graph Theory and Statistical Physics to Real-World Applications},
  ISBN = {9783030941062},
  url = {http://dx.doi.org/10.1007/978-3-030-94106-2},
  DOI = {10.1007/978-3-030-94106-2},
  publisher = {Springer International Publishing},
  author = {Barthelemy,  Marc},
  year = {2022}
}

@article{Barthlemy2003,
  title = {Crossover from scale-free to spatial networks},
  volume = {63},
  ISSN = {1286-4854},
  url = {http://dx.doi.org/10.1209/epl/i2003-00600-6},
  DOI = {10.1209/epl/i2003-00600-6},
  number = {6},
  journal = {Europhysics Letters (EPL)},
  publisher = {IOP Publishing},
  author = {Barthélemy,  M},
  year = {2003},
  month = sep,
  pages = {915–921}
}

@article{Kaiser2004,
  title = {Spatial growth of real-world networks},
  author = {Kaiser, Marcus and Hilgetag, Claus C.},
  journal = {Phys. Rev. E},
  volume = {69},
  issue = {3},
  pages = {036103},
  numpages = {5},
  year = {2004},
  month = mar,
  publisher = {American Physical Society},
  doi = {10.1103/PhysRevE.69.036103},
  url = {https://link.aps.org/doi/10.1103/PhysRevE.69.036103}
}

@article{Maass2007,
  title = {Computational Aspects of Feedback in Neural Circuits},
  volume = {3},
  ISSN = {1553-7358},
  url = {http://dx.doi.org/10.1371/journal.pcbi.0020165},
  DOI = {10.1371/journal.pcbi.0020165},
  number = {1},
  journal = {PLoS Computational Biology},
  publisher = {Public Library of Science (PLoS)},
  author = {Maass,  Wolfgang and Joshi,  Prashant and Sontag,  Eduardo D},
  editor = {Kotter,  Rolf},
  year = {2007},
  month = jan,
  pages = {e165}
}

@misc{Cotteret2024_distributed,
  doi = {10.48550/ARXIV.2405.01305},
  url = {https://arxiv.org/abs/2405.01305},
  author = {Cotteret,  Madison and Greatorex,  Hugh and Renner,  Alpha and Chen,  Junren and Neftci,  Emre and Wu,  Huaqiang and Indiveri,  Giacomo and Ziegler,  Martin and Chicca,  Elisabetta},
  title = {Distributed Representations Enable Robust Multi-Timescale Symbolic Computation in Neuromorphic Hardware},
  publisher = {arXiv},
  year = {2024},
  copyright = {Creative Commons Attribution Share Alike 4.0 International}
}

\end{multicols}
\end{document}